\documentstyle[aps,version2]{revtex}
\begin{document}
\begin{titlepage}
\pagestyle{empty}

\begin{title}
Manifestation of $s\to \Lambda$ fragmentation matrix 
elements via transverse $\Lambda$ polarization 
in  unpolarized $e^-e^+$ annihilation
\end{title}

\author{Wei Lu} 

\begin{instit}
CCAST (World Laboratory), P.O. Box 8730, Beijing 100080, China 
 
and Institute of High Energy Physics, 
 P.O. Box 918(4), Beijing 100039, China\footnote{Mailing address}
 \end{instit}

\author{Xueqian Li} 

\begin{instit}
CCAST (World Laboratory), P.O. Box 8730, Beijing 100080, China 
 
and Department of Physics, Nankai University, 
Tianjin 300071, China 
\end{instit} 

\author{Haiming Hu} 

\begin{instit}
Department of Physics, Peking University, 
Beijing 100871, 300071, China 
\end{instit}

\begin{abstract}
{\small  
Making use of the collinear
expansion technique developed  by Ellis, Furmanski and Petrozio, 
and the special propagator concept invented by Qiu, 
we present a factorization approach  to the photon fragmentation 
tensor for the inclusive Lambda hyperon production. 
As a result, the  structure function  $\tilde F$, 
which is related to the 
transverse polarization of the inclusive 
Lambda  hyperon in unpolarized electron-positron annihilation, 
is expressed  as a combination of four parton fragmentation 
matrix elements. Since the inclusive $\Lambda$ production we considered 
is theoretically  the simplest, 
our study  can be taken as a question to the 
operatability and practability of the first nonleading 
QCD factorization theorem  in the transverse $\Lambda$ polarization 
phenomena.}
\end {abstract}

\end{titlepage} 

\newpage
\pagestyle{plain}
\small
Among the various spectacular  phenomena relating 
to particle spins is 
the  polarization  perpendicular to the production plane
of the  inclusively detected  hyperons in unpolarized  fixed-target 
experiment \cite{hyperon,pondrom}.
Owing to the complexities of  hadron-hadron processes, 
it occurs naturally to one that efforts should be made 
to study  such transverse polarization  phenomena 
 in some simpler and cleaner 
circumstances.  According to the QCD factorization theorem \cite{css}, 
the transverse $\Lambda$ polarizations  in different
processes  share a common set of parton fragmentation 
matrix elements (generalized parton fragmentation functions). 
Hopefully,  one should be able to extract 
the data  about these matrix elements 
in a simpler  chosen process and then make predictions for the 
other complicated inclusive $\Lambda$ productions.  
Such a philosophy is just like the practice that 
people measure the parton distribution functions in the 
deeply inelastic scattering  and thereafter  make predictions 
for other large-momentum-transfer processes.

Presumably, the inclusive $\Lambda$ production by 
unpolarized $e^-e^+$ annihilation  is  the first candidate 
to supply us with information about  transverse-spin-dependent 
fragmentation matrix elements. 
Recently, one of the authors \cite{lw} has pointed out that
due to ``final-state'' interactions,  the inclusively detected spin-half 
fermions in unpolarized $e^-e^+$ annihilation can be 
transversely polarized,  characterized by a structure function $\tilde F$. 
Such a polarization is completely different from  that 
caused by parity violation at the $Z$ resonance \cite{cite1}, 
and easier to measure than  the semi-inclusive $\Lambda\bar\Lambda$
production by unpolarized electron-positron annihilation \cite{cite2}. 
Experimentally, the facilities such as the Beijing Electron-Positron 
Collider (BEPC), the B Factory under construction, 
 and the purposed Tau-Charm Factories provide us a number of 
opportunities to access $\tilde F$ at various 
energy scales.  In this paper, we  present a QCD factorization approach 
to the photon fragmentation tensor for inclusive $\Lambda$ production, 
with the $\Lambda$  spin  perpendicular to the production plane. 
Our results show that $\tilde F$ can be expressed  as a combination 
of  four parton fragmentation matrix elements.  
Unless parton fragmentation matrix elements 
happen to make contributions also in the form of such 
a combination in other inclusive $\Lambda$  production processes, the hope 
is very faint to  establish simple  connections among the transverse
$\Lambda$ polarizations in different reactions. 

 Specifically, what we will consider  is  the process 
\begin{equation}
e^-(p_1) +e^+(p_2) \to \gamma^* (q) \to \Lambda (p,s)+X,
\label{sissy}
\end{equation}
whose invariant  cross section can be written as 
\begin{equation} 
\label{sinn1ex}
E\frac{d\sigma (s)}{d^3 p} 
=\frac{2\alpha^2}{Q^5}L_{\mu\nu}(p_1,q) 
\hat W^{\mu\nu} (q,p,s).
\end{equation} 
At the Born Level,  the leptonic tensor  reads
\begin{eqnarray} 
L^{\mu\nu}(p_1,q)&=&\frac{1}{2}{\rm Tr_D} 
[(\rlap/p_1 +m_e)\gamma^\mu (
\rlap/p_2-m_e)\gamma^\nu] \nonumber \\ 
& = &(-g^{\mu\nu}+ \frac{q^\mu q^\nu}{q^2})
 \frac{q^2}{2}
 -2
(p_1^\mu - \frac{p_1\cdot q}{q^2}q^\mu) 
(p_1^\nu - \frac{p_1\cdot q}{q^2}q^\nu),
\label{lepton}
\end{eqnarray} 
where ${\rm Tr_D}$ signal the trace in the Dirac space. 
Concerning the hadronic tensor, it  is defined as usual 
\begin{equation} 
\hat W_{\mu\nu}(q,p,s)=\frac{1}{4 \pi}
\sum \limits_X \int d^4 \xi \exp ( iq \cdot \xi) 
\langle0| j_\mu (0)|\Lambda (p,s),X\rangle
\langle\Lambda (p,s),X)|j_\nu (\xi)|0\rangle,
\end{equation} 
where $\sum_X$ represents the summation over all the possible 
final states that contain the inclusive hyperon.  The 
electromagnetic  current is defined as $j^\mu=\sum_f 
e_f \bar \psi_f \gamma^\mu \psi_f$, with $f$ being the 
quark flavor index and $e_f$ being the electric charge of the 
quark in unit of the electron charge.  
We will  consider the $\Lambda$ production  via 
the strange quark fragmentation so the flavor index will
be suppressed  wherever possible. 
Taking into account the constraints due to 
the gauge invariance,  hermiticity 
and parity conservation,  we have the following 
general Lorentz decomposition for  $\hat W(q,p,s)$ \cite{lw}: 
\begin{eqnarray} 
\label{dec}
\hat W_{\mu\nu}(q,p,s)&
=&\displaystyle\frac{1}{2}
\left[ (-g_{\mu\nu}+\frac{q_\mu q_\nu}{q^2})\hat F_1(z_B,Q^2)
+(p_\mu -\frac{p\cdot q}{q^2}q_\mu) 
(p_\nu -
\frac{p\cdot q}{q^2}q_\nu)
\frac{\hat F_2(z_B,Q^2)}{p\cdot q} \right]
\nonumber  \\
& & 
+iM \varepsilon_{\mu\nu\lambda\sigma}q^\lambda s^\sigma 
\frac{\hat g_1(z_B,Q^2)}{p\cdot q}
+iM \varepsilon_{\mu\nu\lambda\sigma}q^\lambda 
(s^\sigma -\frac{s\cdot q}{p\cdot q} p^\sigma) 
\frac{\hat g_2(z_B,Q^2)}
{p\cdot q} \nonumber \\
& & 
+M \left[(p_\mu-\displaystyle\frac{p\cdot q}{q^2}
q_\mu)\varepsilon_{\nu \rho\tau\eta}p^\rho q^\tau s^\eta 
+(p_\nu-\displaystyle\frac{p\cdot q}{q^2}q_\nu)
\varepsilon_{\mu \rho\tau\eta}p^\rho q^\tau s^\eta 
\right ]\frac{\tilde F 
(z_B,Q^2)}{(p\cdot q)^2},
\end{eqnarray}
where  $z_B\equiv 2 (p\cdot q)/q^2$, $Q\equiv \sqrt{q^2}$, 
$M$ is the mass of the  $\Lambda$ hyperon,
and $\hat F_1$, $\hat F_2$, $\hat g_1$, 
$\hat g_2$, and $\tilde F$ are the 
scaling structure functions. 
Throughout the  work, we  normalize the spin vector 
in such a way that $s\cdot s =-1$ for the pure state
of a  spin-half particle.

By definition, the transverse $\Lambda$ polarization  reads 
\begin{equation}
P_\Lambda=\frac{
Ed\sigma (s_\uparrow)/d^3 p
-
Ed\sigma (s_\downarrow)/d^3 p 
} 
{
Ed\sigma (s_\uparrow)/d^3 p
+
Ed\sigma (s_\downarrow)/d^3 p 
},
\end{equation}
where $\uparrow$ and $\downarrow$ represent
the $\Lambda$ spin parallel and antiparallel 
to the normal of the  production plane, respectively. 
Substituting eqs. (\ref{lepton}) and (\ref{dec})
into (2), we  will have 
\begin{equation}
P_\Lambda (z_B, Q^2, \theta)= \frac{
4 M{\bf p}^2 \sin\theta\cos\theta\tilde F(z_B,Q^2)  } 
{E\left [2QE \hat F_1(z_B,Q^2) +{\bf p}^2 
\sin^2\theta \hat F_2(z_B,Q^2)\right]  } ,
\label{!}
\end{equation}
where ${\bf p}$ is the momentum  of the inclusively detected 
$\Lambda$ particle  in the center of mass frame
 and  $\theta$  denotes its outgoing angle 
with respect to the  electron beam direction.

Our coordinate system  is specified  by 
letting the $\hat z$ axis  be along the outgoing direction 
of the inclusive hyperon and  putting 
the $\hat x$--$\hat z$ plane in the production plane. 
We adopt the light-cone coordinates 
and parameterize the $\Lambda$ momentum as 
\begin{equation} 
p^\mu=P^\mu +\frac{1}{{2}} M^2 n^\mu, 
\end{equation} 
where 
\begin{equation} 
P^\mu =\frac{1}{\sqrt {2}} ( \sqrt{M^2 +|{\bf p}|^2} +|{\bf p}|)
 (1^+, 0^-, 0_\perp), 
\end{equation} 
\begin{equation} 
 n^\mu =\frac{\sqrt{2}}{ M^2}( \sqrt{M^2 +|{\bf p}|^2} -|{\bf p}|)
(0^+, 1^-, 0_\perp). 
\label{n}
\end{equation} 
Obviously,  $P$ and $n$ are light-like and  they satisfy $P\cdot  n =1$.  
We  will work in the frame in which  $|{\bf p}|$  has a  large value, 
then the plus components of the involved  momenta are dominant.

As can be seen from eq. (\ref{!}),  the transverse 
$\Lambda$ polarization is one-power suppressed, 
i.e., at twist three.  Therefore,  we  will  work  
to the level of twist three in factoring photon fragmentation tensor. 
We choose to work in the light-cone gauge 
specified by  $n\cdot A=0$,  in which the color gauge 
invariance can be  most easily recovered. 
Thus,  our subjects are the diagrams shown  in figs. 1, 2  and 3. 
The main difference between the two diagrams shown in 
fig. 2 and those in fig. 3 lies in that the gluon  exchange 
takes place at the different energy scales. 
 Since the QCD
factorization is an expansion in terms of a hard scattering scale, 
it is necessary to distinguish these  two cases.  In fact, the 
diagrams in fig. 2 has already been discussed by Kane, 
Pumplin and Repko in their classic 1978 paper \cite{Kane}. 
For completeness, we  still include the contributions of these diagrams.

 Two spin-independent structure functions, 
$\hat F_1$ and $\hat F_2$,  have been   thoroughly 
studied in ref. \cite{webber}. According to the 
generalized factorization theorem \cite{QS-fac2}, 
both $\hat F_1$ and $\hat F_2$ are essentially 
at twist two, their first nonleading corrections 
being at twist four.  Since we work up to twist three, 
it is enough to adopt  the twist-two results for $\hat F_1$ and $\hat F_2$. 
That is,  we simply  utilize the  parton model prescription: 
\begin{equation} 
\hat F_1 (z_B)=\frac{2e^2_s }{z_B} \hat f_1 (z_B), 
\end{equation} 
\begin{equation} 
\hat F_2 (z_B)=-\frac{4e^2_s}{ z^2_B} \hat f_1 (z_B), 
\end{equation} 
where $e_s=1/3$ is the  strange quark charge and 
$\hat f_1 (z)$ is the  spin-independent quark fragmentation. 
($\hat f_1(z)$ is also labelled $D(z)$ in the literature, but 
we adopt Jaffe and Ji's notation \cite{JJ,Ji}
 about quark fragmentation functions 
as much as possible.) Henceforth, we  will 
concentrate ourselves  on spin-dependent  structure functions 
and  simply suppress all the spin-independent terms in our presentation. 

To the order at which we work,  
\begin{eqnarray} 
\hat W^{\mu\nu}(q,p,s_\perp)&=& \frac{1}{4\pi N}
\int \frac{d^4 k}{(2\pi)^4} 
{\rm Tr_D} {\rm Tr_C} \left[\left(
H^{\mu\nu}_{(1)}(q,k)
+H^{\mu\nu}_{(2a)}(q,k)
+H^{\mu\nu}_{(2b)}(q,k)
\right) T(k,p,s_\perp)\right]
\nonumber \\ 
& & 
+\frac{1}{4\pi N}\int \frac{d^4 k}{(2\pi)^4} 
\frac{d^4 k_1}{(2\pi)^4} 
{\rm Tr_D} {\rm Tr_C}\left[
H^{\mu\nu\sigma}_{(3a)}(q,k,k_1)
X^\prime _\sigma(k,k_1,p,s_\perp) \right]
\nonumber \\ 
& & 
+\frac{1}{4\pi N}\int \frac{d^4 k}{(2\pi)^4} 
\frac{d^4 k_1}{(2\pi)^4} 
{\rm Tr_D} {\rm Tr_C}\left[
H^{\mu\nu\sigma}_{(3b)}(q,k,k_1)
Y^\prime_\sigma(k,k_1,p,s_\perp)\right],
\label{www}
\end{eqnarray}
where   ${\rm Tr_C}$  represents the trace in the color space,  and 
\begin{equation} 
T_{\alpha\beta}(k,p,s_\perp)= 
\sum\limits_X\int d^4 \xi \exp(-ik\cdot \xi) 
\langle0|\psi_{\alpha}(0)
|\Lambda(p,s_\perp),X\rangle\langle\Lambda(p,s_\perp),X)
|\bar \psi_{\beta}(\xi)|0\rangle, 
\end{equation} 
\begin{eqnarray}
X^{\prime \sigma}_{\alpha\beta} (k_1,k,p,s_\perp)&=& 
\sum\limits_X\int d^4\xi d^4 \xi_1 
\exp (-i(k-k_1)\cdot \xi_1-i k\cdot \xi) 
\nonumber  \\
&  & \times 
\langle0|(-g_s)A^\sigma (\xi_1)\psi_\alpha(0)
|\Lambda (p,s_\perp),X\rangle\langle\Lambda (p,s_\perp),X|
\bar\psi_\beta(\xi) |0\rangle,
\label{u1}
\end{eqnarray} 
\begin{eqnarray}
Y^{\prime \sigma}_{\alpha\beta} (k_1,k,p,s_\perp)&=& 
\sum\limits_X\int d^4\xi d^4 \xi_1 
\exp (-i(k_1-k)\cdot \xi -i k_1\cdot \xi_1) 
\nonumber  \\
&  &\times  \langle0|
\psi_\alpha(0)|\Lambda (p,s_\perp),X\rangle\langle\Lambda (p,s_\perp),X|
\bar\psi_\beta(\xi_1)(-g_s)A^\sigma (\xi) |0\rangle.
\label{v1}\end{eqnarray} 
On writing down eq. (\ref{www}),   we have  utilized 
the shorthand $A^\sigma$ for $A^\sigma_a T_a$, with
$T_a$   the fundamental representation of the color
SU(3) group. For the diagrams in fig. 3, 
the color matrix along with a 
minus strong coupling,  $-g_sT^a_{ij}$,  has been 
isolated from the hard part into the fragmentation matrices. 
Throughout all, the color summation is assumed in our fragmentation 
matrices, so there is a color factor of $1/N$ ($N=3$) in our 
expressions of $\hat W^{\mu\nu}(q,p,s)$.  Furthermore, 
we reserve the subscripts $\alpha$ and $\beta$ for the Dirac indices.

 As Ellis, Petronzio and Furmaski \cite{EFP} have demonstrated 
in the case of the deeply inelastic scattering, 
the leading contributions associated with each diagram 
can be extracted by making an expansion about the 
components of  the parton momenta  that are collinear 
to the corresponding hadron momentum.   Since we 
work up to twist three,  it is needed  to expand the 
lowest-order  diagram to the first derivative term. For this
purpose, we decompose the  strange quark momentum into its 
plus, transverse, and minus components 
\begin{equation}   \label{suda}
k^\mu=\frac{1}{z}P^\mu +k_T^\mu +\frac{k^2-k^2_T}{2k\cdot n}n^\mu.
\end{equation}
In the collinear expansion, the minus component is irrelevant. Therefore, 
we have 
\begin{equation} 
H_{\mu\nu} (q,k)= H_{\mu\nu} (q,P/z)
+ \frac{\partial H_{\mu\nu} (q,k) }{\partial  k^\sigma}
|_{k=P/z} (k-P/z)^\sigma
+\cdots~ . 
\label{B-ex}
\end{equation}
Concerning those diagrams in fig. 3,  whose  leading contributions
come about at twist three,  we can take their  leading terms  
in the collinear expansion.  As a result, we obtain 
\begin{eqnarray} 
\hat W^{\mu\nu}(q,p,s_\perp)&=& \frac{1}{4\pi N} 
\int \frac{dz}{z} 
{\rm Tr_D} {\rm Tr_C} \left[\left(
H^{\mu\nu}_{(1)}(q,P/z)
+H^{\mu\nu}_{(2a)}(q,P/z)
+H^{\mu\nu}_{(2b)}(q,P/z)
\right)T(z,p,s_\perp) \right]
\nonumber \\ 
& & 
+{\rm first~ derivative~ term~ of~the~lowest~order~ diagram~}
\nonumber \\ 
& & 
+\frac{1}{4\pi N}\int d(\frac{1}{z_1}) dz 
{\rm Tr_D} {\rm Tr_C}\left[
H^{\mu\nu\sigma}_{(3a)}(q,P/z,P/z_1)
X^\prime_\sigma(z,z_1,p,s_\perp) \right] 
\nonumber \\ 
& & 
+\frac{1}{4\pi N}
\int d(\frac{1}{z_1}) dz 
{\rm Tr_D} {\rm Tr_C}\left[
H^{\mu\nu\sigma}_{(3b)}(q,P/z,P/z_1)
Y^\prime_\sigma(z,z_1,p,s_\perp)\right], 
\label{18}
\end{eqnarray}
where 
\begin{equation} 
T_{\alpha\beta} (z,p,s_\perp)= 
z \sum\limits_X\int  \frac{d\lambda }{2 \pi}
\exp ( -i\lambda/z) 
\langle0|\psi_{\alpha} (0)|\Lambda (p,s_\perp),X\rangle
\langle\Lambda (p,s_\perp),X|\bar \psi_{\beta} (\lambda n )|0\rangle,
\label{m1}
\end{equation} 
\begin{eqnarray} 
X^{\prime\sigma}_{\alpha\beta} (z_1,z,p,s_\perp)&=& 
\sum\limits_X\int \frac{ d\lambda_1d   \lambda }{(2\pi)^2} 
\exp (-i\lambda_1 (1/z-1/z_1)-i\lambda/z) \nonumber  \\
&  & \times 
\langle0| (-g_s)A^\sigma (\lambda_1   n)\psi_\alpha (0)
|\Lambda (p,s_\perp),X\rangle\langle\Lambda (p,s_\perp),X|
\bar\psi_\beta (\lambda   n) |0\rangle,
\label{xxx}\end{eqnarray} 

\begin{eqnarray}
Y^{\prime\sigma}_{\alpha\beta} (z_1,z,p,s_\perp)&=& 
\sum\limits_X\int \frac{ d\lambda_1d   \lambda }{(2\pi)^2} 
\exp (-i\lambda (1/z_1-1/z)-i\lambda_1/z_1) \nonumber  \\
&  &\times  
\langle0|
\psi_\alpha (0)|\Lambda (p,s_\perp),X\rangle\langle\Lambda (p,s_\perp),X|
\bar\psi_\beta (\lambda_1   n) (-g_s)A^\sigma 
(\lambda   n)|0\rangle. 
\label{yyy}\end{eqnarray} 
Obviously, both $X^{\prime\sigma}_{\alpha\beta} (z_1,z,p,s_\perp)$
and $Y^{\prime\sigma}_{\alpha\beta} (z_1,z,p,s_\perp)$  are not gauge 
invariant objects.  To recover the color gauge invariance, we have two 
options \cite{plb}:  (1) Adding a  derivative 
operator  to $(-g_s) A^\sigma$ to 
form a covariant derivative; (2)  Exploiting integration by parts to 
transform the gluonic field 
$A^\sigma$ into the gluonic  field tensor  $F^{+\sigma}$.

Recalling  that the uncut antiquark propagator 
in the hard part  has the $i\varepsilon$ prescription, we need to 
distinguish two cases.  In the case that the $i\varepsilon$ 
prescription is irrelevant,  one can show by the Ward identity 
\begin{equation}
\frac{\partial H_{\mu\nu} (q,k) }{\partial  k^\sigma}|_{k=P/z} 
=H^{\mu\nu}_{(3a)\sigma} (q, P/z, P/z)
=H^{\mu\nu}_{(3b)\sigma} (q, P/z, P/z), \label{ward}
\end{equation}
then the contributions associated with the  derivative term 
in eq. (\ref{18}) can be combined with leading contributions
of the two diagrams in fig. 3. The net effect of such 
a combination   \cite{qiu} is  to replace $(-g_s) A^\sigma$ 
by the covariant derivative operator 
$D^\sigma=i\partial^\sigma -g_s A^\sigma$ 
in  $X^{\prime\sigma}_{\alpha\beta} (z_1,z,p,s_\perp)$
and $Y^{\prime\sigma}_{\alpha\beta} (z_1,z,p,s_\perp)$. 

 As regard with  the pole contribution in the higher-order diagram, 
the above strategy to reserve the gauge invariance is obviously
not applicable.  This stimulates us to  consider the second 
option stated above, namely,  transforming the gluonic field into  the 
corresponding field tensor.  Here we note that 
in the definitions of  fragmentation matrices,
an $i\epsilon$ prescription is implicit in the exponentials, 
which warranties  the  vanishing 
of matrix elements as $\lambda_1 \to 0$ or $\lambda \to 0$. 
Hence, we can write out 
\begin{eqnarray} 
X^{\prime\sigma}_{\alpha\beta} (z_1,z,p,s_\perp)&=& 
\frac{-ig_s}{1/z-1/z_1+i\varepsilon}
X^{\sigma}_{\alpha\beta} (z_1,z,p,s_\perp), 
\label{xu}\end{eqnarray} 

\begin{eqnarray} 
Y^{\prime\sigma}_{\alpha\beta} (z_1,z,p,s_\perp)&=& 
\frac{-ig_s}{1/z_1-1/z+i\varepsilon}
Y^{\sigma}_{\alpha\beta} (z_1,z,p,s_\perp), 
\label{yv}\end{eqnarray} 
with
\begin{eqnarray} 
X^{\sigma}_{\alpha\beta} (z_1,z,p,s_\perp)&=& 
\sum\limits_X\int \frac{ d\lambda_1d   \lambda }{(2\pi)^2} 
\exp (-i\lambda_1 (1/z-1/z_1)-i\lambda/z) \nonumber  \\
&  & \times 
\langle0| n^-F^{+\sigma} (\lambda_1   n)\psi_\alpha (0)
|\Lambda (p,s_\perp),X\rangle\langle\Lambda (p,s_\perp),X|
\bar\psi_\beta (\lambda   n) |0\rangle,
\label{uu0}\end{eqnarray} 

\begin{eqnarray}
Y^{\sigma}_{\alpha\beta} (z_1,z,p,s_\perp)&=& 
\sum\limits_X\int \frac{ d\lambda_1d   \lambda }{(2\pi)^2} 
\exp (-i\lambda (1/z_1-1/z)-i\lambda_1/z_1) \nonumber  \\
&  &\times  
\langle0|
\psi_\alpha (0)|\Lambda (p,s_\perp),X\rangle\langle\Lambda (p,s_\perp),X|
\bar\psi_\beta (\lambda_1   n) n^-F^{+\sigma}
(\lambda   n)|0\rangle. 
\label{vv0}\end{eqnarray} 

However, the collinear expansion  scheme is not 
a satisfactory procedure to extract 
nonleading  twist  contributions. As Qiu has demonstrated  \cite{qiu}, 
the leading term in the collinear 
expansion contains as well the nonleading contributions. 
A  remedy to the Ellis-Furmanski-Petronzio  procedure 
is the ``special'' propagator technique  invented
by Qiu. Now  we explain what Qiu's special propagator \cite{qiu} is. 
Consider  the  quark propagator 
in fig. 3 that links the electromagnetic vertex to the 
quark-gluon one. From our parameterization (\ref{suda}), 
one can write 
\begin{equation} 
\frac{i\rlap/k}{k^2+i\varepsilon}=
\frac{i\rlap/{\hat k}}{k^2+i\varepsilon }
+\frac{i\rlap/n}{2k\cdot n}.
\label{special}\end{equation} 
where 
\begin{equation} 
\hat k^\mu=\frac{1}{z} P^\mu +k^\mu_T -\frac{k^2_T}{2k\cdot n} n^\mu
\end{equation} 
is the on-shell part  of $k^\mu$. 
Physically,  the $  i n\cdot \gamma /(2k\cdot n)$  piece 
describes a ``contact'' interaction in the 
light-cone, so it should be included into the 
hard  partonic interaction part. 
Since $  i n\cdot \gamma /(2k\cdot n)$  is a part of the propagator, 
Qiu termed  it the special propagator.  
Graphically,  it is labelled by adding a 
bar on the normal propagator. 
In our  twist-three case,   only 
$one$  such special propagator  along with the 
connected quark-gluon vertex  needs to be invoked, 
either on the  left-hand side or on the right-hand side 
of the final-state cut.   In other words,  twist-three 
contributions hidden in the leading term of the  collinear 
expansion of the lowest-order diagram  can 
be taken into account by  including the two diagrams in fig. 4. 
Considering that the diagrams shown in fig. 4 can be obtained 
from the lowest-order diagram by pulling out a 
quark propagator and its connected quark-gluon vertex, 
it can also  be stated that the non-contact part of its linking 
propagator should be  discarded, 
because they have been included in $T_{\alpha\beta}
(k,p,s_\perp)$. It should be stressed that Qiu's  
special propagator prescription can naturally 
reserve the  gauge invariance
for the hard part \cite{qiu}.

From the above discussions, we know that 
the formula  for calculating the hadronic tensor reads 
\begin{eqnarray} 
\hat W^{\mu\nu}(q,p,s_\perp)&=& \frac{1}{4\pi N} 
\int \frac{dz}{z} 
{\rm Tr_D} {\rm Tr_C} \left[\left(
H^{\mu\nu}_{(1)}(q,P/z)
+H^{\mu\nu}_{(2a)}(q,P/z)
+H^{\mu\nu}_{(2b)}(q,P/z)
\right)T(z,p,s_\perp)\right]
\nonumber \\ 
& & 
+\frac{1}{4\pi N}\int d(\frac{1}{z_1}) dz 
{\rm Tr_D} {\rm Tr_C}\left[\left(
H^{\mu\nu\sigma}_{(3a)}(q,P/z,P/z_1)
+
H^{\mu\nu\sigma}_{(4a);{\rm spec}}(q,P/z)
\right)U_\sigma(z,z_1,p,s_\perp) \right] 
\nonumber \\ 
& & 
+\frac{1}{4\pi N}
\int d(\frac{1}{z_1}) dz 
{\rm Tr_D} {\rm Tr_C}\left[\left(
H^{\mu\nu\sigma}_{(3b)}(q,P/z,P/z_1)
+H^{\mu\nu\sigma}_{(4b);{\rm spec}}(q,P/z)
\right)V_\sigma(z,z_1,p,s_\perp)\right]
\nonumber \\ 
& & 
+\frac{1}{4\pi N}\int d(\frac{1}{z_1}) dz  \frac{-ig_s}
{1/z- 1/z_1+i\varepsilon}
 \nonumber \\ 
& & \times 
{\rm Tr_D} {\rm Tr_C}\left[\left(
H^{\mu\nu\sigma}_{(3a)}(q,P/z,P/z_1)
+
H^{\mu\nu\sigma}_{(4a);{\rm spec}}(q,P/z)
\right)X_\sigma(z,z_1,p,s_\perp) \right] 
\nonumber \\ 
& & 
+\frac{1}{4\pi N}
\int d(\frac{1}{z_1})dz \frac{-ig_s}{1/z_1- 1/z+i\varepsilon}
\nonumber \\ 
& & \times {\rm Tr_D} {\rm Tr_C}\left[\left(
H^{\mu\nu\sigma}_{(3b)}(q,P/z,P/z_1)
+H^{\mu\nu\sigma}_{(4b);{\rm spec}}(q,P/z)
\right)Y_\sigma(z,z_1,p,s_\perp)\right], 
\label{www1}
\end{eqnarray}                  
where 
\begin{eqnarray} 
U^\sigma_{\alpha\beta} (z_1,z,p,s_\perp)&=& 
\sum\limits_X\int \frac{ d\lambda_1d   \lambda }{(2\pi)^2} 
\exp (-i\lambda_1 (1/z-1/z_1)-i\lambda/z) \nonumber  \\
&  & \times 
\langle0| {D}^\sigma (\lambda_1   n)\psi_\alpha (0)
|\Lambda (p,s_\perp),X\rangle\langle\Lambda (p,s_\perp),X|
\bar\psi_\beta (\lambda   n) |0\rangle,
\label{u}\end{eqnarray} 

\begin{eqnarray}
V^\sigma_{\alpha\beta} (z_1,z,p,s_\perp)&=& 
\sum\limits_X\int \frac{ d\lambda_1d   \lambda }{(2\pi)^2} 
\exp (-i\lambda (1/z_1-1/z)-i\lambda_1/z_1) \nonumber  \\
&  &\times  
\langle0|
\psi_\alpha (0)|\Lambda (p,s_\perp),X\rangle\langle\Lambda (p,s_\perp),X|
\bar\psi_\beta (\lambda_1   n) \stackrel{\leftarrow}{D^\sigma}
(\lambda   n)|0\rangle. 
\label{v}\end{eqnarray} 
with $\stackrel{\leftarrow}{D^\sigma}=-i\stackrel{\leftarrow}{\partial^\sigma}
 -g_sA^\sigma$.

 Since the gluonic field can either be converted into a field 
tensor or combined with a derivative operator to form a covariant 
derivative operator, there  seems to be a mixing 
between $X^\sigma_{\alpha\beta}(z_1,z,p,s_\perp)$  and 
$U^\sigma_{\alpha\beta}(z_1,z,p,s_\perp)$, and 
$Y^\sigma_{\alpha\beta}(z_1,z,p,s_\perp)$  and 
$V^\sigma_{\alpha\beta}(z_1,z,p,s_\perp)$. 
In other words, one may question  if the  higher-twist
contributions in eq. (\ref{www1}) are double-counted. 
Fortunately, this does not cause  any  problem in our 
present case.  Our computational processes shows that 
the twist-three matrix elements with a covariant 
derivative operator are irrelevant 
of the pole part of the higher-order diagrams, 
whereas those with a  gluonic field tensor make contributions 
only  by  virtue of the pole  term.

 To arrive at  factorized expressions,  we decompose 
the above  three  fragmentation matrices in the 
Dirac and Lorentz  spaces. 
We suppress  the  spin-independent terms 
and go to the twist-three level
for those spin-dependent terms.  For $T_{\alpha\beta}(z,p,s_\perp)$, 
we have 
\begin{eqnarray} 
T_{\alpha\beta}(z,p,s_\perp)
&=& \hat h_1 (z) (\gamma_5 \rlap/s_\perp \rlap/P)_{\alpha\beta} 
+M\hat g_T (z) (\gamma_5 \rlap/s_\perp)_{\alpha\beta} 
+M\tilde g_T(z)\varepsilon^{\delta\eta\rho\tau}s_{\perp\eta}
P_\rho n_\tau (\gamma_\delta)_{\alpha\beta}
+\cdots. \label{dec1}
\end{eqnarray} 
As for the decomposition of our   two-variable  fragmentation matrices, 
we note  that  the gluonic field tensor is  related to the covariant 
derivative operator via  
$F^{\mu\nu}= 1/(ig_s)[D^\mu, D^\nu]$. Therefore,  they  have 
different behaviours under the time-reversal transformation. 
In discussing  the spin asymmetry, it is more convenient to 
employ the adjoint parity-time-reversal transformation instead of the 
individual parity or time-reversal transformation. 
Under the adjoint transformation,  the covariant derivative 
is even  while the  filed tensor behaves  odd.  As a result, 
we have the following decompositions 

\begin{eqnarray} 
U^\sigma_{\alpha\beta}(z_1,z,p,s_\perp)&=& 
\frac{iM}{2z}\hat G_1 (z_1,z) 
\varepsilon_{\sigma \rho\tau\eta}s^\rho_\perp P^\tau n^\eta
\rlap/P_{\alpha\beta} 
+
\frac{M}{2z}\hat G_2(z_1,z) s^\sigma_\perp (\gamma_5 \rlap/P)_{\alpha\beta} 
\nonumber \\ 
& & 
+\frac{M}{2z}\tilde G_1 (z_1,z) 
\varepsilon_{\sigma \rho\tau\eta}s^\rho_\perp P^\tau n^\eta
\rlap/P_{\alpha\beta} 
+
\frac{iM}{2z}\tilde G_2(z_1,z) s^\sigma_\perp (\gamma_5 \rlap/P)_{\alpha\beta} 
+\cdots ,
\label{udec}
\end{eqnarray}

\begin{eqnarray} 
V^\sigma_{\alpha\beta} (z_1,z,p,s_\perp)&=&
 -\frac{iM}{2z}\hat G_1 (z_1,z) 
\varepsilon_{\sigma \rho\tau\eta}s^\rho_\perp P^\tau n^\eta
\rlap/P_{\alpha\beta} 
+
\frac{M}{2z}\hat G_2 (z_1,z) s^\sigma_\perp (\gamma_5 \rlap/P)_{\alpha\beta}
\nonumber \\ 
& & 
 +\frac{M}{2z}\tilde G_1 (z_1,z) 
\varepsilon_{\sigma \rho\tau\eta}s^\rho_\perp P^\tau n^\eta
\rlap/P_{\alpha\beta} 
-
\frac{iM}{2z}\tilde G_2 (z_1,z) s^\sigma_\perp (\gamma_5 \rlap/P)_{\alpha\beta}
+\cdots ,
\label{vdec}
\end{eqnarray} 

\begin{eqnarray} 
X^\sigma_{\alpha\beta}(z_1,z,p,s_\perp)&=& 
\frac{M}{4\pi z }\hat H_1 (z_1,z) 
\varepsilon_{\sigma \rho\tau\eta}s^\rho_\perp P^\tau n^\eta
\rlap/P_{\alpha\beta} 
+
\frac{iM}{4\pi z }
\hat H_2(z_1,z)  s^\sigma_\perp (\gamma_5 \rlap/P)_{\alpha\beta} 
\nonumber \\ 
& & 
+\frac{iM}{4\pi z}\tilde H_1 (z_1,z) 
\varepsilon_{\sigma \rho\tau\eta}s^\rho_\perp P^\tau n^\eta
\rlap/P_{\alpha\beta} 
+
\frac{M}{4\pi z}\tilde H_2(z_1,z) 
s^\sigma_\perp (\gamma_5 \rlap/P)_{\alpha\beta} 
+\cdots ,
\label{xdec}
\end{eqnarray}

\begin{eqnarray} 
Y^\sigma_{\alpha\beta} (z_1,z,p,s_\perp)&=&
\frac{M}{4\pi z}\hat H_1 (z_1,z) 
\varepsilon_{\sigma \rho\tau\eta}s^\rho_\perp P^\tau n^\eta
\rlap/P_{\alpha\beta} 
-
\frac{iM}{4\pi z}\hat H_2 (z_1,z)s^\sigma_\perp (\gamma_5 \rlap/P)_{\alpha\beta}
\nonumber \\ 
& & 
 -\frac{iM}{4\pi z}\tilde H_1 (z_1,z) 
\varepsilon_{\sigma \rho\tau\eta}s^\rho_\perp P^\tau n^\eta
\rlap/P_{\alpha\beta} 
+
\frac{M}{4\pi z}\tilde H_2 (z_1,z) 
s^\sigma_\perp (\gamma_5 \rlap/P)_{\alpha\beta}
+\cdots .
\label{ydec}
\end{eqnarray} 
In our work,  we use  a 
tilde  to signal those  matrix elements
that arise from the hadronic final-state interactions.
The definitions of the matrix elements can be 
easily obtained by  projecting the corresponding matrix 
elements with  the appropriate projectors. 
$\hat h_1 (z) $, $\hat g_T(z) $,  and  $\tilde g_T(z)$ 
(labelled $\hat g_{\bar T}(z)$ by Jaffe and Ji)
are Jaffe and Ji's  single-variable  quark fragmentation 
functions \cite{JJ,Ji}. $\hat G_1(z_1, z) $ and $\hat G_2(z_1, z) $ 
were also introduced first   in ref. \cite{Ji}, 
but   the other  six  two-variable twist-three  fragmentation 
matrix elements appear first time in this paper. 
These eight  constitute a complete set of transverse-spin-dependent, 
two-variable, and  twist-three  parton fragmentation matrix elements.

Substituting eqs. (\ref{dec1})--(\ref{ydec}) into  (\ref{www1}), 
we obtain  after some algebra 
\begin{eqnarray} 
\hat W_{\mu\nu}(q,p,s_\perp)&=& 
\frac{ie^2_s }{ z_0^2 (P\cdot q)}
\left[z_0 m_s \hat h_1(z_0)+ M \hat g_T(z_0)\right]
\varepsilon_{\mu\nu\tau\rho}q^\tau s^\rho_\perp.
\nonumber \\ 
& &  +\frac{e^2_s}{4 z^2_0  (P\cdot q)^2 } 
\left[ 
\frac{N^2-1}{N} \alpha_s m_s \hat h_1(z_0)
-\frac{1}{2}M g_s z_0  \left(\hat H_1(z_0,z_0) -\hat H_2(z_0,z_0)\right)
-M\tilde g_T (z_0)\right] 
\nonumber \\ 
& &  \times 
\left[(p_\mu-\displaystyle\frac{p\cdot q}{q^2}
q_\mu)\varepsilon_{\nu \rho\tau\eta}p^\rho q^\tau s^\eta_\perp
+(p_\nu-\displaystyle\frac{p\cdot q}{q^2}q_\nu)
\varepsilon_{\mu \rho\tau\eta}p^\rho q^\tau s^\eta_\perp
\right ],\label{res}
\end{eqnarray} 
where $z_0\equiv 2(P\cdot q)/q^2$. According to eqs. (8-10),
$z_0$ is same as $z_B$ up to  a piece of 
twist-four effects.

Here some  notes are in order.

(1). To arrive at this 
expression with manifest  electromagnetic gauge invariance, 
we  have employed the following projection relations: 
\begin{equation} 
\int d(\frac{1}{z_1}) [ \hat G_1 (z_1,z) + \hat G_2 (z_1,z)] = 
-\frac{1}{z}\hat g_T (z)+ \frac{m_s}{M} \hat h_1 (z),  \label{proj1}
\end{equation} 
\begin{equation} 
\int d(\frac{1}{z_1}) [ \tilde G_1 (z_1,z) + \tilde G_2 (z_1,z)] = 
-\frac{1}{z}\tilde g_T (z).  \label{proj2}
\end{equation} 
 The relations of this 
type was first identified by Effremov and Teryaev \cite{ET} in 
their  studies of  single transverse spin asymmetries.  Equation 
(\ref{proj1}),  with its quark mass term neglected,   has ever
appeared in ref. \cite{Ji}. In this quark fragmentation version,  
we included the effects of  the quark mass  
associated with chiral-odd, 
twist-two quark fragmentation function $\hat h_1(z).$  However, 
there is no quark mass term in eq. (\ref{proj2}), because there
is no final-state-interaction-caused quark fragmentation 
function at twist two. 

(2). The diagrams in fig. 2 can  be taken as vertex  corrections to 
the lowest-order diagram in fig. 1.  In fact, their contributions  
have been discussed in the  framework of the quark parton model
in ref. \cite{Kane}. At the  twist at which we work,  
the contributions of those two diagrams 
in fig. 2 can be included by replacing  the fundamental 
electromagnetic vertex $\gamma^\mu$  in fig. 1 with the effective
vertex operator
\begin{equation} 
\Gamma^\mu=\gamma^\mu  \left[1+f(q^2)\right]+ \frac{1}{2m_s} 
\sigma^{\mu\tau}q_\tau g(q^2), 
\end{equation} 
or similarly for $\gamma^\nu$.  For the  transverse polarization  
requires flipping the chirality, only the Dirac tensor  part is 
relevant. By considering the imaginary part of the decay amplitudes
for $\gamma^\ast\to s \bar s$,  one can obtain  \cite{soviet}, 
\begin{equation}                                                  
g(q^2)=\frac{N^2-1}{2} \frac{e^2_s \alpha_s m^2_s}
{\sqrt{q^2 (q^2-4m^2_s)}}.
\end{equation} 
Obviously, the contributions of  the vertex corrections are  
insignificant for they vanish in the scaling limit.

(3).  Our factorization result, eq. (\ref{res}) is complete 
to the  $O(Q^{-1})$. Therefore, we can make substitutions 
$P \to p$ and $z_0 \to z_B$, which includes some effects 
at two higher twist. Then, we can confront our result 
with  the general decomposition of $\hat W^{\mu\nu}(q,p,s)$, 
eq. (\ref{dec}).   As a result, 
\begin{equation} 
\hat g_1(x_B) +\hat g_2(x_B) =\frac{1}{9 z^2_B}
\left[\frac{m_s}{M}z_B  \hat h_1(z_B)+ \hat g_T(z_B)\right], \label{cao1}
\end{equation} 
\begin{equation} 
\tilde F(z_B)=\frac{1}{36 z^2_B} 
\left[ 
\frac{N^2-1}{N} \frac{m_s}{M} \alpha_s  \hat h_1(z_0)
-\frac{1}{2} g_s z_B  \left(\hat H_1(z_B,z_B) -\hat H_2(z_B,z_B)\right)
-\tilde g_T (z_B)\right] . \label{cao2}
\end{equation} 

In summary, our study shows 
that structure function $\tilde F$ is  related to 
a combination of four parton fragmentation matrix elements, even we 
assume  that the $\Lambda$ particle production is predominantly 
via the strange quark fragmentation.  At present, there is no reliable 
way  to calculate them from the first principle--QCD.  Actually, 
the predictive power of QCD factorization lies in that  different
processes are related  by a common set of nonperturbative matrix 
elements.  Although $\tilde F$ can be measured by  experiments, 
it   cannot be expected to extract easily information about individual 
fragmentation matrix elements.  If one can justify that 
one of them  plays a leading role in certain phase space, however, 
it will become a possible practice. At the time being, such
justifications is obviously beyond reach.  At first sight, one 
may argue that the contribution associated with $\hat h_1(z)$, 
which  characterizes the  chiral symmetry breaking effects 
due to the  strang quark mass, is  negligible. But  it is 
still unknown to what extent the other three  matrix elements, 
$\tilde g_T(z)$, $\hat H_1(z,z)$ and $\hat H_2(z,z)$, are
important.  Therefore, this work exemplified the   
potential difficulties in applying the generalized QCD 
factorization theorem to the transverse $\Lambda$ polarization 
problem,  that is, the number of the  relevant  fragmentation 
matrix elements  is so large that  one cannot find 
a  simple standard process to  normalize the others. 
Nevertheless, 
if  the parton  fragmentation matrix elements  happen to contribute 
in the combinational form  that we obtained,  it is still 
possible to establish a simple  connection between the  various 
transverse $\Lambda$ polarization phenomena.  Although such a hope 
is very faint, we feel it is still desirable to  work on a 
factorized expression for  transverse $\Lambda$
 polarization in hadron-hadron 
process by applying the scheme outlined in this paper. 

One of the  authors (W. L.) thanks  X. Ji   for 
correspondences as well as for  a prepublication 
copy of ref. \cite{Ji}.

\newpage
\pagestyle{empty}

\centerline{Figure Captions}

\begin{enumerate}

\item{The lowest-order cut diagram for the inclusive 
$\Lambda$ hyperon production by a time-like photon.}

\item{The cut diagram for the inclusive 
hyperon production by a time-like photon
with one  gluon correlation before the quark fragmentation.} 

\item{The cut diagram for the inclusive 
hyperon production by a time-like photon
with one  gluon correlation at the  quark fragmentation stage.}

\item{The cut diagram for the inclusive 
$\Lambda$ hyperon production by a time-like photon
with  one  gluon radiation in the quark fragmentation. 
One ``special'' propagator  is pulled down into the hard partonic interaction 
part, either (a) on the left-hand side or (b) on the right-hand side 
of the final-state cut.} 

\end{enumerate} 

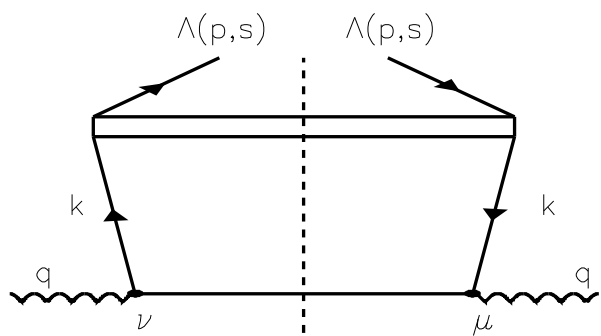
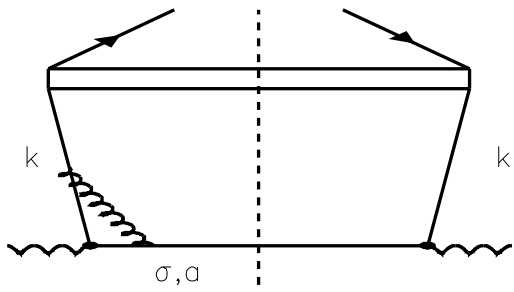
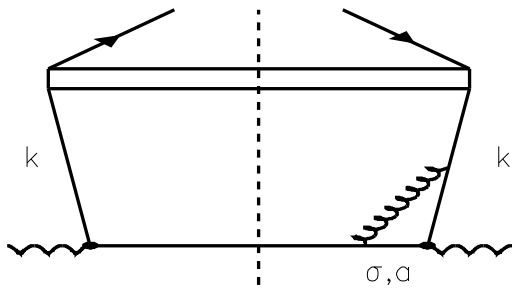
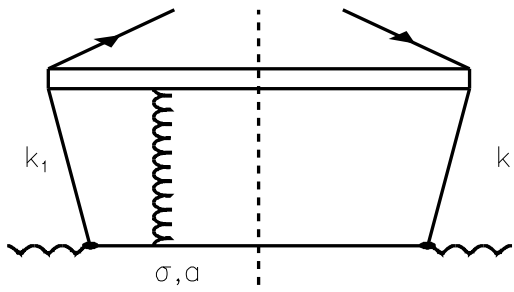
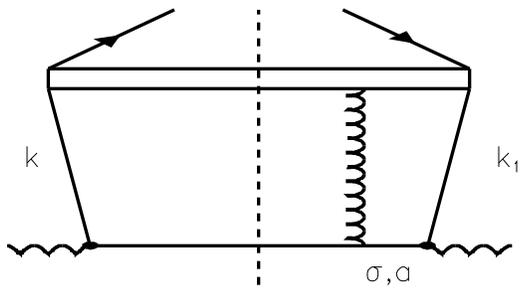
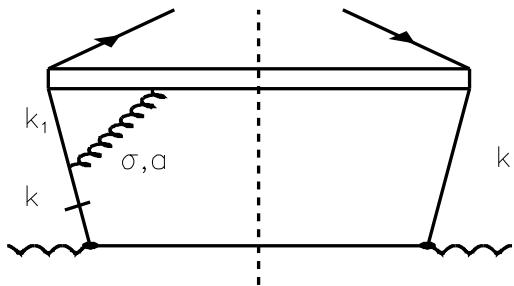
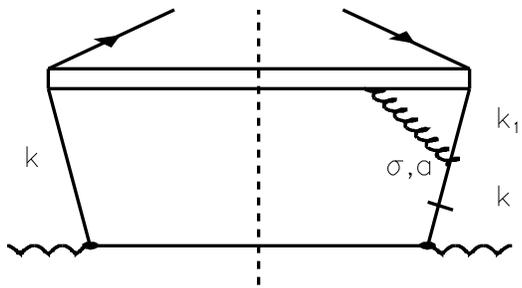
\end{document}